**A contrastive learning foundation model based on perfectly aligned sample pairs for remote sensing images**

Hengtong Shen, Haiyan Gu, Haitao Li, Yi Yang and Agen Qiu

State Key Laboratory of Spatial Datum, Chinese Academy of Surveying and Mapping, Beijing, China

**ABSTRACT**

Self-supervised learning (SSL) facilitates the pre-training of foundation models without reliance on costly labeled data. Among SSL methods, Contrastive learning (CL) excels at extracting robust semantic representations, even in the presence of complex interference. However, despite the success of CL in general computer vision, the significant domain gap necessitates specific adaptations for remote sensing imagery. To this end, we present a novel self-supervised method called PerA, which produces all-purpose remote sensing features through semantically perfectly aligned sample pairs. Specifically, PerA obtains features from sampled views by applying spatially disjoint masks to augmented images rather than random cropping. Our framework learns high-quality representations by ensuring consistency between teacher and student networks and predicting learnable mask tokens. Compared to previous contrastive methods, our method demonstrates higher memory efficiency and supports larger batch sizes due to the processing of sparse inputs. Additionally, the proposed method exhibits remarkable adaptability to uncurated remote sensing data and mitigates the impact of the potential semantic inconsistency. We also collected an unlabeled pre-training dataset, which contains about 5 million remote sensing images. We conducted experiments on multiple downstream task datasets and achieved performance comparable to previous state-of-the-art methods with a limited model scale, validating the effectiveness of our approach. We hope this work will contribute to practical remote sensing image interpretation.

**CONTACT** Haiyan Gu 🖂 guhy@casm.ac.cn

## 1. Introduction

In recent years, deep learning approaches have increasingly become the primary means for ground surface interpretation tasks such as natural resource surveys (Pei et al. 2021; Chen et al. 2018; Chang et al. 2024), crop monitoring (Nguyen et al. 2020; Hong et al. 2024), and ecological environment protection (Yuan et al. 2020; Du et al. 2024). However, relying on manually labeled remote sensing (RS) data, it is difficult to match the dataset scale to increasingly large models. To address this issue, many self-supervised learning (SSL) methods have been employed to leverage the vast amount of unlabeled remote sensing data available (Reed et al. 2023; Stojnic and Risojevic 2021). Numerous works have significantly reduced the reliance on labeled data by leveraging self-supervised or unsupervised learning techniques (Lv et al. 2025; Yang J et al. 2025; Yang S et al. 2025). Using well-designed pretext tasks, SSL methods produce pseudo-labels to learn general and intelligible representations (Balestriero et al. 2023). Among the two main categories of SSL, generative and contrastive, contrastive learning (CL) is more competitive in image linear classification (Chen, Xie, and He 2021). By distinguishing between similar and dissimilar data points, CL methods learn global-level semantics from large amounts of data and significantly enhance the model's representational capabilities. Although numerous works have explored applying CL methods to remote sensing images, these methods are typically designed for natural images, which differ significantly from remote sensing images. Specifically, these differences can be summarized as follows:

(1) RS images are captured by various sensors, resulting in a significant domain distance between them and natural images.

(2) RS images are taken from a bird's eye view, which causes a large gap in spatial resolution compared to natural images. Many different categories of objects in RS images are usually small and scattered.

(3) RS image samples are usually uncurated, as they are obtained by cropping large-size remote sensing imagery. Unlike most natural image datasets, these wild samples are neither object-centered nor category-balanced.

These properties of remote sensing images make it difficult to transfer good features from models pre-trained on natural image datasets like ImageNet (Wang et al. 2022). Besides, most CL methods are based on instance discrimination and similar pretext tasks, which consider augmented images as different views (crops) and encourage similarity between them. Such a pretext task generally works well in most cases, and the mild semantic inconsistencies it introduces can enhance the model's robustness. However, in certain special cases, as illustrated in Figure 1(a) and (b), improper cropping may lead to intense semantic inconsistencies, especially in remote sensing images (Muhtar, Zhang, and Xiao 2022; Song et al. 2023). After random cropping, there is a high probability that the operation removes semantic information that is beneficial for pretraining, or generates sample pairs whose semantics are entirely unrelated. These limitations call for more efficient and RS-adapted methods to develop a better feature extractor with unlabeled remote sensing data.

To achieve more precise semantics, many studies have started examining the effective underlying mechanisms of contrastive learning approaches and understanding how they extract features from unlabeled data (Wang and Isola 2020; Ji et al. 2023). A reasonable interpretation is that, with much stronger data augmentations, CL methods attempt to find correlations between sparse signals and separate them from spurious dense noise. Since different coordinate of dense noise are independent to each other, we can decorrelate the dense noise by randomly applying two completely opposite masks to images that learn better representations as much as possible (Wen and Li 2021). Moreover, with random masks, these positive pairs come from the same original image, making them semantically aligned but inconsistent in appearance. As shown in Figure 1 (c) and (d), randomly masked samples rarely result in semantic inconsistencies, even in uncurated remote sensing images containing tiny objects. As the size of the mask units (patch size) decreases, the probability of generating semantically inconsistent sample pairs decreases as well, since objects in remote sensing images are less likely to be masked completely. Once the patch size is reduced to be smaller than the smallest object in the image, the semantic information between sample pairs will be identical.

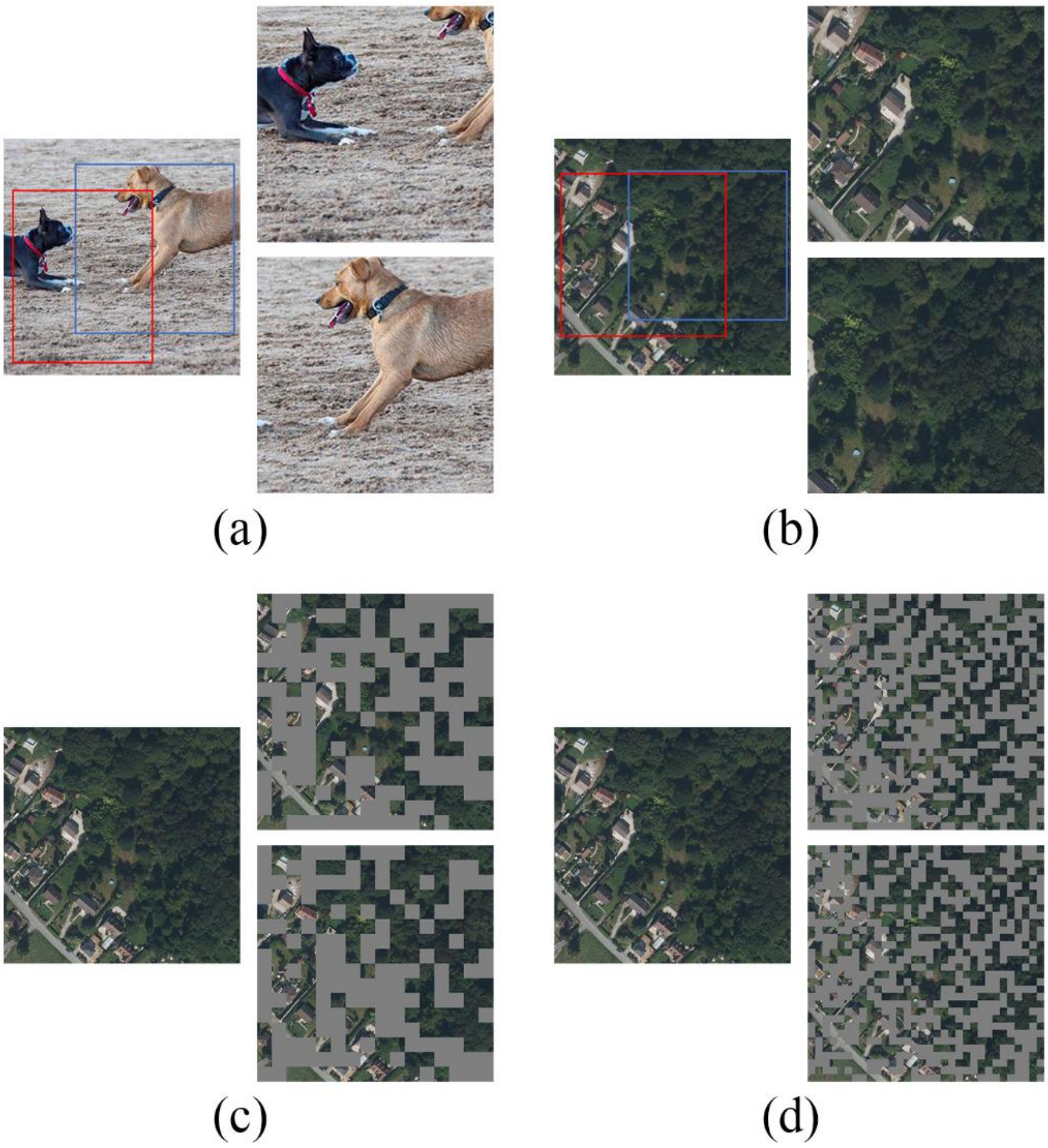

Figure 1. Comparison of random cropped views on ImageNet and random masked views on uncurated RS images. (a) Semantic misalignment in random cropped views on ImageNet; (b) Semantic misalignment in random cropped views on uncurated RS images; (c) Semantic alignment preserved in 32 × 32 patch random masked views on uncurated RS images; (d) Semantic alignment preserved in 16 × 16 patch random masked views on uncurated RS images.

Numerous works have found that CL methods learn better global representations by capturing relations between positive sample pairs, resulting in better performance in global classification tasks (Muhtar et al. 2023). However, they sometimes lack ability in dense prediction tasks like object detection (Li, Xie, Chen, et al. 2021). In contrast, Masked image modeling (MIM) methods, which learn representations by reconstructing the missing information or features from the masked image, are usually less suited for linear classification, but exhibit robust dense prediction capabilities (Zhou et al. 2022; Huang et al. 2023). From a certain point of view, CL and MIM methods complement each other. Additionally, the reconstructing ability of MIM benefits sparse input, which is likely to be masked partly. These motivate us to propose a new remote sensing pre-training framework that accepts sparse image inputs and combines CL and MIM methods. In this study, we present PerA, a simple but effective remote sensing contrastive learning method that processes sparse sample pairs which are perfectly aligned in semantics. PerA fully leverages the advantages of CL, MIM, and sparse semantic information, so that it can adapt to most common uncurated RGB remote sensing images. To ensure that the model has sufficient training data to learn a general and effective representation, we designed a data processing pipeline to obtain unlabeled remote sensing images for pre-training and collected a remote sensing pre-training dataset, which contains about 5 million unlabeled remote sensing images. remote sensing images in the dataset were captured from satellite maps, ranging in spatial resolution from 2 to 10 m. Thanks to the global land cover products of high accuracy, these images cover six continents and are balanced in categories. Based on the pre-training dataset and PerA, we trained a remote sensing foundation model with unlabeled images, which can obtain features with accurate semantic information and can be widely used in a variety of downstream tasks. We evaluate our model on the AID dataset (Xia et al. 2017), ISPRS Potsdam dataset (Sherrah 2016), and LEVIR-CD (Chen and Shi 2020) dataset, etc., achieving competitive results in every downstream task.

In conclusion, our main contributions can be summarized as follows:

(1) We developed an automatic pipeline for capturing unlabeled pre-training data. By random sampling, we obtained around 5 million unlabeled remote sensing images globally and compiled them into the dataset RSRSD-5m. To the best of our knowledge, this is one of the largest publicly available unlabeled remote sensing datasets.

(2) We designed a novel SSL method called PerA, which is tailored to the characteristics of remote sensing images. It demonstrates that sparse sample pairs with perfectly aligned semantics improve transfer performance and representation capabilities.

(3) Extensive experiments across various downstream tasks, including image classification, semantic segmentation, and change detection, show that our pre-training framework achieves comparable to or even surpasses previous state-of-the-art results, which verify the superiority of our method.

## 2. Related work

In this section, we mainly introduce relevant works of contrastive learning methods in RS and computer vision (CV). Finally, we discuss the currently available methods used to resolve semantic inconsistencies in CL.

### *2.1. Contrastive learning methods in CV*

The idea of contrastive learning, which encourages similarity between semantic features, was first introduced in Bromley et al. (1993). However, it was not until recent significant works that CL achieved groundbreaking progress in the field of CV. By discriminating among individual instances, CL methods can learn more general representations (Wu et al. 2018). Nevertheless, early CL methods like CMC (Tian, Krishnan, and Isola 2020) and SimCLR (Chen et al. 2020) struggled with storing vast high-dimensional signals of negative samples. These methods either simply employ huge batch sizes, which consume a tremendous amount of computational resources, or rely on a memory bank, which can cause representation irrelevance. These issues were not solved until MoCo (He et al. 2020) was proposed. MoCo utilizes a queue and a moving-averaged encoder to build a large and consistent dictionary of sample pairs. The memory-efficient queue provides sufficient negative samples while the slowly updated moving-averaged encoder ensures that representations come from similar extractors. MoCo successfully suggested that CL methods can still reach competitive results with unlabeled data only.

To store representations of negative samples, these methods designed different frameworks to get a large dictionary with extensive semantic information. However, BYOL (Grill et al. 2020), an approach that simply adds a linear predictor head to the classic CL structure, demonstrates that CL methods can learn meaningful information only with positive sample pairs. As long as the model collapse is prevented, CL methods can learn visual representations without negative samples, saving memory and reducing computational complexity significantly. So far, these advantages have benefited many Vision Transformer-based CL methods like IBOT (Zhou et al. 2021) and DINO (Caron et al. 2021), which exhibit superior representational ability through global attention, scalability, and flexibility.

### *2.2. Contrastive learning methods in RS*

SSL methods make it possible to learn remote sensing representations without manually annotated labels, which consume a significant amount of cost. Although MIM methods, which resemble autoencoders in architecture, have been widely used to learn various remote sensing representations, CL methods have been widely utilized in recent years due to their flexibility and global semantic learning capabilities. Stojnic and Risojevic et al. (2021) applied remote sensing images to the CMC (Tian, Krishnan, and Isola 2020). They demonstrated that even though the pre-training dataset was smaller than ImageNet, the self-supervised model trained on RGB remote sensing images outperformed the supervised model trained on ImageNet in remote sensing downstream tasks. Ayush et al. (2021) and Manas et al. (2021) utilized time-series remote sensing images as positive sample pairs, effectively leveraging the natural properties of time-series images for contrastive learning. However, they did not account for the potential degradation in representation quality due to land cover category changes over time. Scheibenreif et al. (2022) and Prexl and Schmitt (2023) utilized medium and low-resolution remote sensing data, which are very effective for large-scale regional observations. However, they are not adequate for more general high-resolution remote sensing surface monitoring tasks. The flexibility of CL makes it easier to take advantage of the diversity of remote sensing data to train multimodal models by combining multiple modalities, including visible, multispectral, hyperspectral, SAR, LiDAR, and so on. Jain, Schoen-Phelan, and Ross (2022) and Wang, Braham, Xiong, et al. (2023) utilized sample pairs of multispectral and SAR images to obtain representations, while Duan et al. (2022) and Wang, Gao, Dong, et al. (2023) incorporated hyperspectral data and LiDAR point cloud data for pre-training. Feng et al. (2023) proposed a method that utilizes data distributed across multiple modalities, such as visible, SAR, hyperspectral, and near-infrared. The various data modalities are modeled independently in parallel and then fused into a single backbone to obtain better results. Although these works demonstrate the adaptability of CL methods for remote sensing data in different types, these data are expensive and difficult to obtain. In contrast, RGB remote sensing data remains the most common and available.

### *2.3. Semantic alignment*

Contrastive learning methods attempt to reach maximum agreement between differently augmented sample pairs. To learn semantic representations from distorted samples, CL methods tend to filter out noise introduced by augmentations and find meaningful signals in high-level features. However, positive sample pairs produced by random cropping, trying to capture features in shared regions, may sometimes have no intersection at all. Therefore, these positive sample pairs, which are semantically inconsistent, can only obtain signals at low levels. Many studies have employed various methods to improve the existing pretext tasks to obtain semantically richer and more consistent positive sample pairs. LeOCLR (Alkhalefi, Leontidis, and Zhong 2024) attempts to prevent inconsistencies by using the original image as an anchor. SauMoCo (Kang et al. 2020) and CDL (Jung et al. 2021) simply treat neighboring cropped images as pairs. Peng et al. (2022) and Mishra et al. (2021) designed additional modules which generate proposal regions so that sample pairs are bounded within them. IndexNet (Muhtar, Zhang, and Xiao 2022) and PixPro (Xie et al. 2021) attempt to learn representations through scoring mechanisms to evaluate similarity at the pixel level. Self-EMD (Liu, Li, and Sun 2020) and SCFS (Song et al. 2023) utilize well-designed metrics to encourage features to learn close semantics. These improvements either fail to fully leverage the invariant semantic features in the samples or incorporate complex computations. To address these issues, we designed a straightforward method that can save memory and reach maximum semantic alignment with disjoint masks.

## 3. Methodology

In this section, we first present the pipeline for obtaining remote sensing image tiles of various spatial resolutions to collect a label-free pre-training remote sensing dataset. Then, we describe the overall approach in detail, including disjoint masks, the pre-training method, and the pixel prediction module.

### *3.1. Pre-training dataset*

In order to achieve better generalization of the model, we designed an automatic pipeline to build an unlabeled remote sensing pre-training dataset from internet platforms. Employing Google Earth Engine, we first collect random sample points across six continents, which are demonstrated in the schematic Figure 2. In Europe, Asia, and North America, where urban areas are much larger, the number of sampling points is twice that of other continents. Sampling was conducted with consideration of category boundaries defined by the ESRI 10-m Land Use/Land Cover map (Karra et al. 2021), ensuring a diverse representation of various land cover types. Due to the significant demand for urban monitoring, we allocated a sampling rate of 60% for urban regions and 10% each for farmland, wetlands, forests, and watersheds. In addition, we collected 1000 sample points in each of the four regions: desert, ice, grassland, and cloud to ensure diversity. With category-balanced sample points, we download remote sensing image tiles from online map engines, Google Maps and Bing Maps. Ranging in spatial resolution from 0.3 to 10 m, these tiles have been standardized to 512 × 512. Finally, we built an unlabeled pre- training remote sensing dataset with about 5 million image slices, which is called Random Sampled Remote Sensing Dataset 5 m (RSRSD-5m). To the best of our knowledge, this is one of the largest publicly available unlabeled remote sensing datasets.

Due to random sampling, slices in RSRSD-5m are quite similar to the remote sensing images used for practical monitoring applications in industry. Unlike curated datasets that are pinpointed and filtered to obtain objects, our dataset is barely processed but rich in semantic information and data volume. We randomly sampled 100 thousand images from both well-curated dataset TOV-RS (Tao et al. 2023) and ours, and evaluated them on the AID dataset. In Section 4.3.1, we validate the effectiveness of the dataset collected by our method through comparative experiments, which indicate that even without curation, the model achieves performance comparable to that trained on curated datasets. Additionally, during formal pre-training, we incorporate the TOV dataset and the Million-AID dataset (Long et al. 2021) to obtain more comprehensive representations in higher resolution, which are detailed in Table 1.

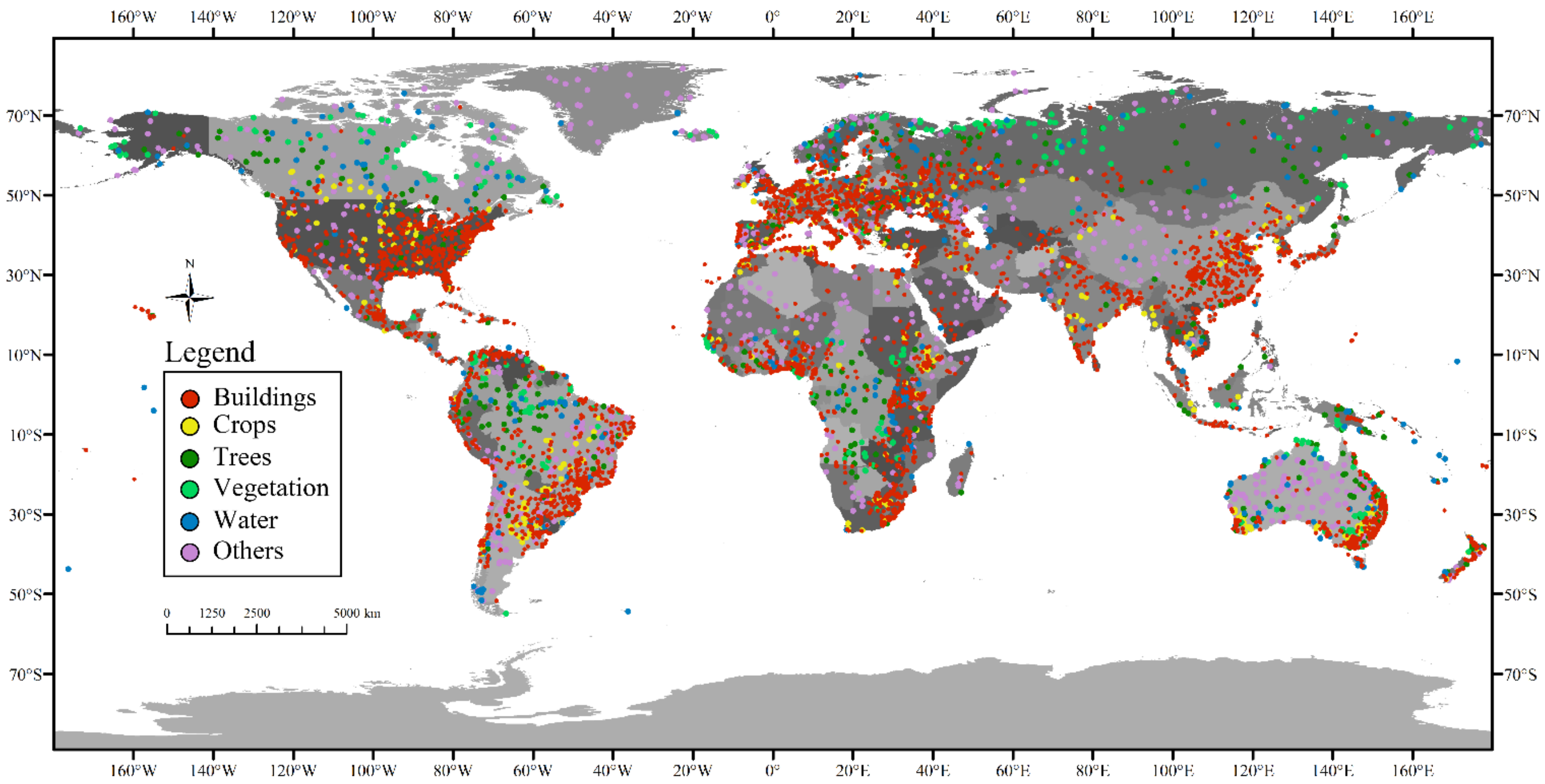

Figure 2. Illustration of global random sample points of the RSRSD-5m dataset.

Table 1 Comparison among different large high-resolution remote sensing image datasets.

| Dataset | Num. of images | Spatial resolution (m) | Source | Image sizes |
|---|---|---|---|---|
| MLRSNet (2020) | 109,161 | 0.1 to 10 | Google Earth | 256 |
| Levir-KR (2021) | 1,431,950 | 0.8 to 16 | Gaofen-1, Gaofen-2, Gaofen-6 | 256 |
| RSD46-WHU (2017) | 117,000 | 0.5 to 2.4 | World Map, Google Maps | 256 |
| fMoW (2018) | 132,716 | - | DigitalGlobe | - |
| Million-AID (2021) | 1,000,848 | 0.5 to 153 | Google Earth | 256/512 |
| TOV-RS (2023) | 1,088,941 | 1 to 20 | Google Earth | 600 |
| **RSRSD-5m** | **4,686,059** | **0.3 to 10** | **Google Maps, Bing Maps** | **512** |

### *3.2. PerA*

The overview framework of PerA is depicted in Figure 3. Our approach is designed to make all efforts to generate sample pairs that are aligned in semantics and inconsistent in appearance. To this end, we proposed an improved method that is adapted to remote sensing, based on DINOv2 (Oquab et al. 2023). In this section, we will illustrate our method in detail in disjoint masks, the pre-training method, and pixel prediction.

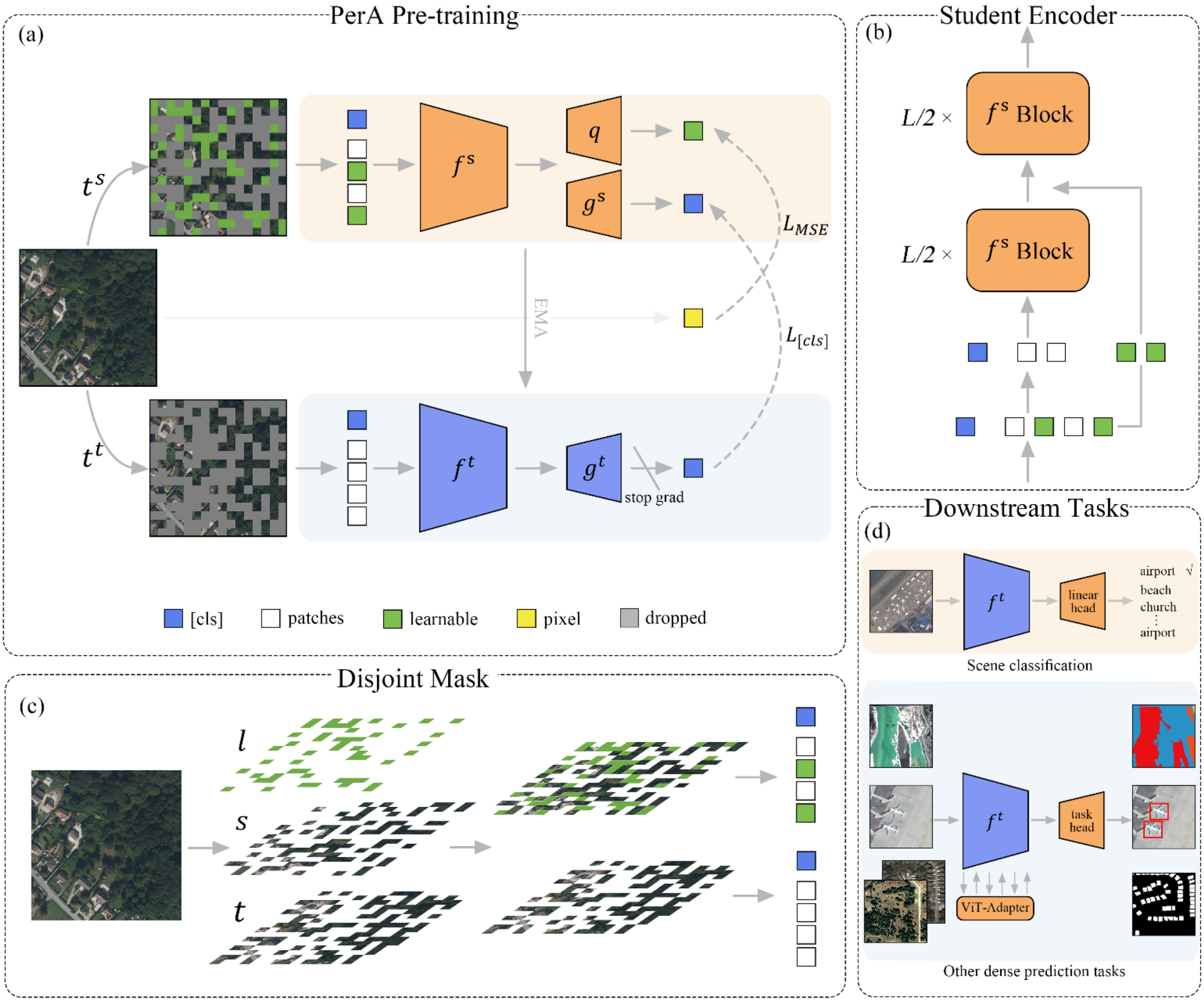

Figure 3. Overview framework of PerA: (a) pre-training; (b) student encoder structure; (c) disjoint mask; (d) fine-tuning on downstream interpretation tasks.

#### *3.2.1. Perfectly aligned sample pairs*

We utilize disjoint masks to obtain sparse inputs to achieve more precise and concise semantics. Specifically, as shown in Figure 3 (c), the random masks $\boldsymbol{M}_l$, $\boldsymbol{M}_s$ and $\boldsymbol{M}_t$ partition the image into three spatially disjoint subsets: $l$, $s$, and $t$, which represent the learnable region, the student input region, and the teacher input region, respectively. The typical ratio for this partition is 2:2:6. Given an arbitrary image sample $\boldsymbol{x} \sim D$ from the data distribution, $\boldsymbol{x}$ is embedded into a sequence of fixed-size patches. To construct the student input embedding $\boldsymbol{x}^s$, the class token $cls$ (blue), the learnable mask tokens corresponding to subset $l$ (green), and the student region input patches $s$ (white) are concatenated in their original sequence. Conversely, the teacher input embedding $\boldsymbol{x}^t$ consists solely of the class token and the teacher input patches $t$. Mathematically, this process can be formulated as:

$$\boldsymbol{1} = \boldsymbol{M}_l + \boldsymbol{M}_s + \boldsymbol{M}_t \tag{1}$$

$$\boldsymbol{x}^s = cls \oplus (\boldsymbol{M}_s \odot \boldsymbol{x}) \oplus (\boldsymbol{M}_l \odot \boldsymbol{x}_{\text{mask}}) \tag{2}$$

$$\boldsymbol{x}^t = cls \oplus (\boldsymbol{M}_t \odot \boldsymbol{x}) \tag{3}$$

where $\boldsymbol{x}_{\text{mask}}$ denotes the matrix of learnable mask tokens, $\boldsymbol{1}$ denotes the all-ones matrix, $\odot$ denotes the element-wise multiplication, and $\oplus$ denotes concatenation preserving the original spatial order of the patches.

### *3.2.2. PerA pre-training*

As shown in Figure 3 (a), PerA consists of a student network and a teacher network, both networks utilize the Vision Transformer (ViT; Dosovitskiy et al. 2020) as a backbone network. The student contains three parts: student encoder $f^s$, projector $g^s$, and predictor $q$. The teacher, which has the same architecture as student only without a predictor, contains teacher encoder $f^t$ and projector $g^t$. Given a sampled image $\boldsymbol{x} \sim D$, we firstly apply two different augmentations $t^s$ and $t^t$ which contain spatial augmentations, color distortions, and disjoint masks detailed in Section 3.2.1. The spatial augmentations in $t^s$ and $t^t$, including random flipping and random resized cropping, have the same parameters to ensure spatial alignment. We follow DINOv2's color distortion strategy, but additionally apply random grayscale. After patch embedding, we apply patchwise disjoint masks to augmented views and obtain sparse inputs $\boldsymbol{x}^s = t^s(\boldsymbol{x})$ and $\boldsymbol{x}^t = t^t(\boldsymbol{x})$. Containing $[cls]$ tokens $x^s_{[cls]}$, patch tokens $\boldsymbol{x}^s_p$ and learnable mask tokens $\boldsymbol{x}^s_{\text{mask}}$, the student input $\boldsymbol{x}^s$ will be processed separately as shown in Figure 3 (b). The learnable mask tokens $\boldsymbol{x}^s_{\text{mask}}$ will participate in encoding halfway, and patch tokens $\boldsymbol{x}^s_p$ will be discarded. After encoding, the $x^s_{[cls]}$ will be processed by projector $g^s$ to obtain projection $\hat{x}^s_{[cls]}$. The predictor $q$ will convert learnable mask tokens $\boldsymbol{x}^s_{\text{mask}}$ to mask prediction $\hat{\boldsymbol{x}}^s_{\text{mask}}$. Similarly, the teacher will input $x^t$ consisting of $x^t_{[cls]}$ and $\boldsymbol{x}^t_p$, and output a projection $\hat{x}^t_{[cls]}$. By predicting pixel value of masked patches, the $\hat{\boldsymbol{x}}^s_{\text{mask}}$ is encouraged to reduce mean squared error (MSE) between original images. For each image $p_i$ in mini batch, we have:

$$L_{\text{MSE}} = \frac{1}{B}\sum_{i=1}^{B}\left(\hat{\boldsymbol{x}}^s_{i,mask} - \boldsymbol{p}_i\right)^2 \tag{4}$$

where $\hat{\boldsymbol{x}}^s_{i,mask}$ is the predicted value corresponding to each $\boldsymbol{p}_i$. The loss function used to compute the loss for $L_{[cls]}$ can be expressed as:

$$L_{[cls]} = -H\left(\frac{\boldsymbol{t} - c}{tpt_t}\right)\log\left(H\left(\frac{\boldsymbol{s}}{tpt_s}\right)\right) \tag{5}$$

where $\boldsymbol{t}$ and $\boldsymbol{s}$ denote the output features of two networks, and the $tpt_s$ and $tpt_t$ denote the temperature values that control the sharpness of the distribution of the stent and teacher outputs. $H(\cdot)$ can be formulated as:

$$H(\cdot) = \frac{\exp(\cdot)}{\sum_{k=1}^{K}\exp(\cdot)} \tag{6}$$

Given a momentum $n$, $c$ updates with the batch mean of the teacher's output to reduce the probability of model collapse, which the formula can express as:

$$c \longleftarrow nc + (1-n)\frac{\sum_{b=1}^{B}\hat{\boldsymbol{x}}^t_{b,[cls]}}{B} \tag{7}$$

where $B$ denotes the batch size and $\hat{\boldsymbol{x}}^t_{b,[cls]}$ denotes every teacher output projection in the batch. Additionally, we incorporate the Koleo loss to encourage dispersion of the output feature vectors in the embedding space and prevent potential model collapse. The formula for the KoLeo loss is as follows:

$$L_{\text{KoLeo}} = -\frac{1}{B}\sum_{i=1}^{B}\log(\|\boldsymbol{z_i} - \boldsymbol{z}'_{\boldsymbol{i}}\|_2 + \varepsilon) \tag{8}$$

where $B$ denotes the batch size, $\|\boldsymbol{z_i} - \boldsymbol{z}'_{\boldsymbol{i}}\|_2$ denotes the euclidean distance between output features and $\varepsilon$ is a constant. The student network updates the weights by backpropagating, following the weighted-sum loss function:

$$L = C_1 L_{[cls]} + C_2 L_{\text{MSE}} + C_3 L_{\text{Koleo}} \tag{9}$$

where $C_1$, $C_2$, and $C_3$ are weighting constants, empirically assigned values of 1.0, 0.1, and 0.01, respectively.

Distinct from the student, the teacher updates using an exponential moving average (EMA) with momentum $m$, which can be delineated as:

$$\boldsymbol{\theta_t} \leftarrow m\boldsymbol{\theta_t} + (1-m)\boldsymbol{\theta_s} \tag{10}$$

where $\boldsymbol{\theta_s}$ and $\boldsymbol{\theta_t}$ denote student and teacher weights. The momentum $m$ is usually large, ranging in [0, 1] to ensure that the teacher weights $\boldsymbol{\theta_t}$ can be updated slowly.

### *3.2.3. Downstream transferring*

The transfer learning technique helps the model to achieve better performance in the target domain by transferring the knowledge learned by pre-training. Aiming to learn a representation that can easily transfer to various downstream tasks, we employ the teacher encoder as the backbone and concatenate it with task-specific heads. However, the Vision Transformer is trained on sequential data and lacks the spatial priors inherent in convolutional neural networks. This limitation makes ViT perform poorly on dense prediction tasks. To address this issue, we utilize ViT-Adapter (Chen et al. 2022) to enhance performance on dense prediction tasks, such as segmentation. As shown in Figure 3 (d), we use the original teacher encoder in classification, whereas for dense prediction tasks, we warp the encoder with ViT-Adapter to capture more spatial information.

## 4. Experiments

To evaluate the performance of our proposed PerA, we first introduce our pre-training implementation setting in detail. Next, we conduct experiments on scene classification, semantic segmentation, and change detection. At the end of the section, we present detailed ablation experiments and feature visualizations. All experiments were implemented with NVIDIA A100 40G GPUs.

### *4.1. Pre-training implementation*

We utilized ViT as the backbone and pre-trained the model on our unlabeled pre-training dataset, which was described in detail in Section 3.1. By default, we employed the AdamW optimizer (Loshchilov and Hutter 2017) with a batch size of 192. We conducted pre-training on a slim ViT-G/16-1024 model with an embedding dimension of 1024 and 505 M parameters for 200 epochs. The learning rate is linearly ramped up during the first 20 epochs to its base value and then decayed with a cosine schedule after warmup. The weight decay and the momentum $m$ also follow the cosine schedule, but slowly increase throughout the pre-training. The student temperature is set to a constant value of 0.1. In contrast, the teacher temperature grows linearly to a maximum value in the first 30 epochs and remains constant until the end of pre-training. We utilized a large number of head prototypes, specifically 131,072, and a high drop path rate of 0.3. The student and teacher mask ratios were set to 0.3 and 0.5, with the remaining 0.2 allocated to the learnable portion. For image augmentations, we adopted multi-crop, random flipping, random resized cropping, followed by color distortion consisting of a random sequence of brightness, contrast, saturation, hue adjustments, and a grayscale conversion.

### *4.2. Downstream tasks performance*

We employed the encoder-decoder paradigm to adapt the learned representations to a specific task, as explained in Section 3.2.3. We evaluate our model on the AID dataset (Xia et al. 2017), UC Merced dataset (Yang and Newsam 2010), ISPRS Potsdam dataset (Sherrah 2016), LoveDA dataset (Wang et al. 2021), LEVIR-CD dataset (Chen and Shi 2020), and DIOR dataset (Li et al. 2020), which are representative remote sensing datasets for scene classification, semantic segmentation, change detection, and object detection, respectively. The results indicated that our model achieved performance comparable to previous state-of-the-art methods with a limited model scale, demonstrating the effectiveness and generalization capability of our method.

#### *4.2.1. Scene classification*

For scene classification, we evaluate our model on the AID dataset and the UC Merced dataset. We first introduce these datasets in detail:

AID dataset: AID dataset is a large-scale dataset for aerial scene classification. Consisting of 10,000 high-resolution aerial images ranging in 0.5 to 8 m, these images are standardized to a size of 600 × 600 and are divided into 30 categories. The dataset is designed to have high intra-class diversity and low inter-class dissimilarity, making classification more challenging.

UC Merced dataset: UC Merced scene classification dataset (UCM) is a standard dataset widely used in remote sensing image scene classification, which contains 2100 high-resolution remote sensing images. These image tiles consist of 21 land-use classes derived from aerial orthoimagery with a pixel resolution of one foot. All these images are downloaded from the United States Geological Survey (USGS) National Map of 20 regions of the United States. In each class, the dataset contains 100 images of size 256 × 256, and all of them were carefully selected to prevent homogeneity.

Implementation details: We utilized a linear classifier to transfer our pre-trained model to the domain of the dataset. For each dataset, all the images are resized to 512 × 512 before scene classification. For the AID dataset, we divide the dataset into training and test sets in a ratio of 2 to 8 according to Sun et al. (2022). We finetuned the model with AdamW optimizer and Cross Entropy Loss for 200 epochs. The batch size was set to 16, and weight decay was set to 5e-6. The learning rate was ramped up to its maximum value 1e-4, during the initial 20 epochs, and then it was gradually decreased using a cosine scheduler. For the UCM dataset, we employed an 8:2 training-test split based on Sun et al. (2022) and a 5:5 split based on Wang et al. (2022). Unlike the AID experiments, we set the learning rate and the weight decay to 5e-5 and 1e-4. We set the drop path rate to 0.3 to enhance the model's generalization ability. For data augmentations, we used random flipping, random cropping, and color distortion.

Results and analyses: We evaluate our pre-trained model's capabilities through fine-tuning performance and report overall accuracy (OA). As shown in Table 2, we compared the finetuning performance with other remote sensing foundation models on the AID dataset. We note that our model can reach the best level with a limited model scale, even though we used only 20% of the data as the training set, which proves the effectiveness of our method. On the other hand, as shown in Table 3, under the commonly used 80% training data setting, our model achieved up to 100% overall accuracy on the test set of the UCM dataset. Although such results are partly influenced by randomness and do not prove that the finetuned model can fully fit the entire dataset, they still demonstrate the strong scene classification capability of our pre-trained model. Given the relatively low complexity of the task setting, we additionally conducted experiments utilizing 50% training data and compared our model with several other existing remote sensing pre-training methods. Our model still achieved the best performance, that demonstrate the superiority of our proposed method.

Table 2 Comparisons with other RS foundation models on the AID dataset.

| Method | Arch. | OA (%) |
|---|---|---|
| RSP (2022) | ViTAEv2-S | 96.91 |
| RVSA (2022) | ViTAE-B | 97.03 |
| Scale-MAE (2023) | ViT-L | 96.44 |
| SeCo (2021) | ResNet-50 | 93.47 |
| SatMAE (2022) | ViT-L | 95.02 |
| TOV (2023) | ResNet-50 | 95.16 |
| CMID (2023) | Swin-B | 96.11 |
| GASSL (2021) | ResNet-50 | 93.55 |
| SSL4EO (2023) | ViT-S | 91.06 |
| CACo (2023) | ResNet-50 | 90.88 |
| SatLas (2023) | Swin-B | 94.96 |
| GFM (2023) | ViT-L | 95.47 |
| Skysense (2024) | ViT-L + Swin-H | 97.68 |
| Ringmo (2022) | Swin-B | 96.90 |
| PerA | ViT-G/16-1024 | 97.13 |

Table 3 Comparisons with other RS foundation models on the UC Merced dataset , where TR indicates training set ratio.

| Method | OA (%) | |
|---|---|---|
| | TR = 50% | TR = 80% |
| RSP (2022) | 99.62 | 99.90 |
| GeoKR (2021) | 95.49 | - |
| GeCo (2022) | 96.32 | - |
| SeCo (2021) | - | 97.86 |
| CMID (2023) | - | 99.48 |
| RVSA (2022) | 99.70 | - |
| Ringmo (2022) | - | 99.84 |
| PerA | 99.71 | 100.0 |

#### ***4.2.2. Semantic segmentation***

We conducted experiments on the ISPRS Potsdam dataset and the LoveDA dataset, which are remote sensing datasets for semantic segmentation widely utilized to evaluate the dense prediction ability of models. The datasets are introduced as follows:

ISPRS Potsdam dataset: The Potsdam dataset contains 38 high-resolution aerial images within 0.5 m, and all of them are 6000 × 6000 pixels in size. The dataset was obtained in Potsdam city, Germany, which contains large building blocks, narrow streets, and dense settlement structures. The dataset has six categories of labels in the dataset, including impervious surfaces, building, low vegetation, tree, car, and clutter. The dataset also provides near-infrared band imagery and digital surface models (DSMs). However, we only utilized RGB optical data for the segmentation task.

LoveDA dataset: The LoveDA dataset consists of 2522 training, 1669 validation, and 1796 test images. The dataset was obtained from Nanjing, Changzhou, and Wuhan in China through the Google Earth platform. Each image has a fixed pixel size of 1024 × 1024 and 0.3 spatial resolution. The dataset includes seven classes: buildings, roads, water, barren land, forests, agriculture, and background. The images are sampled in both urban and rural areas using different data collection strategies to increase their diversity and difficulty.

Implementation details: The models are finetuned with ViT-Adapter to enhance performance on dense prediction tasks as described in Section 3.2.3. The UPerNet (Xiao et al. 2018) was employed to capture multi-level spatial features. We applied its decoder to our pre-trained backbone. In the UPerNet decoder, we set hidden channels to 512 and output channels to 256. For the Potsdam dataset, we separated the 38 images into a training set of 24 images and a test set of 14 images following the setting of previous works. All the images were cropped into 512 × 512, and we overlapped them when it is out of the boundary. We finetuned our model for 200 epochs with AdamW optimizer, Cross Entropy Loss, and batch size of 8. The learning rate would increase from 0 to 1e-4 in 20 initial epochs, and slowly decrease to 5e-6. The drop path rate was set to 0.1, and the weight decay was set to 1e-4. We only adopted random flipping and random resized cropping for augmentations. For the LoveDA dataset, we utilized official splits to finetune our model. Since the ground truth labels of the test set were not provided, we uploaded the inference results to the official evaluation server (https://codalab.lisn.upsaclay.fr/competitions/421). We set the maximum learning rate to 5e-5, while keeping the other settings the same as those used in the experiments on the Potsdam dataset.

Results and analyses: We reported the mean F1-score of experiment results on the Potsdam dataset. As shown in Table

4, compared with other methods, our method can achieve a competitive segmentation capability on the Potsdam dataset. Under the same experiment setting, our model outperformed the majority of existing pre-training methods and achieved less than a 1% difference in the mF1 metric compared to the current state-of-the-art, despite having a significantly smaller model size and pre-training data scale. For the LoveDA dataset, we estimate the performance of our model using mean intersection over union (mIoU). As shown in Table 5, our method achieved a 54.02% mIoU on the test set of the LoveDA dataset. Although it does not surpass the current state-of-the-art result, it still ranks among the top-performing pre-training methods.

Table 4 Comparisons with other RS foundation models on the Potsdam dataset.

| Method | Arch. | mF1 (%) |
|---|---|---|
| RSP (2022) | ViTAEv2-S | 90.64 |
| RVSA (2022) | ViTAE-B | 91.22 |
| Scale-MAE (2023) | ViT-L | 91.54 |
| SeCo (2021) | ResNet-50 | 89.03 |
| SatMAE (2022) | ViT-L | 90.63 |
| TOV (2023) | ResNet-50 | 92.03 |
| CMID (2023) | Swin-B | 91.86 |
| GASSL (2021) | ResNet-50 | 91.27 |
| SSL4EO (2023) | ViT-S | 91.54 |
| CACo (2023) | ResNet-50 | 91.35 |
| SatLas (2023) | Swin-B | 91.28 |
| GFM (2023) | ViT-L | 91.85 |
| Skysense (2024) | ViT-L + Swin-H | 93.99 |
| Ringmo (2022) | Swin-B | 91.27 |
| PerA | ViT-G/16-1024 | 93.08 |

Table 5 Comparisons with other RS foundation models on the LoveDA dataset.

| Method | Arch. | mIoU (%) |
|---|---|---|
| RSP (2022) | ViTAEv2-S | 53.02 |
| SeCo (2021) | ResNet-50 | 44.64 |
| BFM (2023) | ViT-G/12 | 54.40 |
| RVSA (2022) | ViTAE-B | 52.44 |
| CACo (2023) | ResNet-50 | 48.89 |
| TOV (2023) | ResNet-50 | 49.70 |
| GeRSP (2024) | ResNet-50 | 50.56 |
| Geo-Aware (2019) | ResNet-50 | 49.37 |
| MTP (2024) | ViT-L + RVSA | 54.17 |
| SelectiveMAE (2024) | ViT-L | 54.31 |
| GASSL (2021) | ResNet-50 | 48.76 |
| PerA | ViT-G/16-1024 | 54.02 |

#### *4.2.3. Change detection*

Change detection plays a crucial role in land use monitoring, environmental protection, and disaster response. With the growing emphasis on change detection tasks, the ability of remote sensing foundation models to perform change detection has become an important indicator for evaluating their overall capability. For change detection, we adopted LEVIR-CD to evaluate our method. In the following, we provide a detailed introduction to the LEVIR-CD dataset:

LEVIR-CD dataset: The LEVIR-CD dataset, a high-resolution building change detection dataset, consists of 445, 64, and 128 pairs of remote sensing images each in the training, validation, and test sets. The images were captured in Texas, USA, including Austin, Lakeway, Bee Cave, Buda, Kyle, Manor, Pflugerville, Dripping Springs, etc., between 2002 and 2018. Each image is 1024 × 1024 pixels in size. All of the temporal remote sensing images in the dataset are in a spatial resolution of 0.5 m. Additionally, the seasonal changes and the illumination changes are also introduced to the dataset, which makes it more difficult to interpret.

Implementation details: For change detection tasks, we combined our encoder with the BIT head (Chen, Qi, and Shi 2021) and finetuned it for 500 epochs. We retained the original parameters of the BIT model, and simply replaced the ResNets to our pre-trained encoders. Same as general change detection approaches, the backbone performs two forward passes and shares the same weights in both. The LEVIR-CD dataset provides standard training, validation, and test sets. Following common practices in existing methods (Sun et al. 2022), we cropped all image tiles into samples of 512 × 512 pixels. We finetuned our model with pixel-level Cross Entropy Loss and AdamW optimizer. The batch size was set to 8, and the drop path rate was set to 0.2. After the learning rate climbs to 1e-4, it will decrease to 1e-6 following a cosine scheduler. We adopted random flipping and random resized cropping for augmentations and disabled all color jittering.

Results and analyses: We evaluate the change detection capability of our model on the LEVIR-CD dataset, and report the F1 score of the foreground class in the prediction results. The model's performance is available in Table 6, and we notice that our model can achieve a competitive result of 92.34% F1 score. The result outperforms most existing RS pre-trained models and exhibits a marginal gap of only 0.24% compared to the state-of-the-art performance, which demonstrates the superiority of our approach in the change detection task.

Table 6 Comparisons with other RS foundation models on the LEVIR-CD dataset.

| Method | Arch. | F1 (%) |
|---|---|---|
| RSP (2022) | ViTAEv2-S | 90.93 |
| RVSA (2022) | ViTAE-B | 90.86 |
| Scale-MAE (2023) | ViT-L | 92.07 |
| SeCo (2021) | ResNet-50 | 90.14 |
| SatMAE (2022) | ViT-L | 87.65 |
| CMID (2023) | Swin-B | 91.72 |
| GASSL (2021) | ResNet-50 | 78.19 |
| SSL4EO (2023) | ViT-S | 89.05 |
| CACo (2023) | ResNet-50 | 81.04 |
| SatLas (2023) | Swin-B | 90.62 |
| GFM (2023) | ViT-L | 91.73 |
| Skysense (2024) | ViT-L + Swin-H | 92.58 |
| Ringmo (2022) | Swin-B | 91.86 |
| PerA | ViT-G/16-1024 | 92.34 |

#### *4.2.4. Object detection*

Object detection tasks focus on identifying and localizing specific categories of objects, and play a critical role in urban planning, traffic monitoring, maritime surveillance, and disaster response. Unlike natural image object detection, remote sensing object detection tasks often deal with challenges such as varying scales, complex backgrounds, small object sizes, and diverse imaging conditions, making it more challenging for remote sensing interpretation. For object detection, we conducted experiments on extensively utilized DIOR dataset to evaluate the object detection performance of our model. The following is a detailed introduction to the DIOR dataset:

DIOR dataset: The DIOR dataset is a large-scale public remote sensing object detection dataset that contains 20 common remote sensing object categories, including Airplane, Ship, Overpass, Windmill, Chimney, Harbor, Stadium Vehicle, etc. Captured on the Google Earth platform, all the images have a size of 800 × 800 pixels and spatial resolutions ranging from 0.5 to 30 m. The dataset provides official splits of trainval set and testing set that contain 11,725 images and 11,738 images. The objects in dataset have a large range of object size variations, not only in terms of spatial resolutions, but also in the aspect of inter- and intra-class size variability across objects. The diversity and broad coverage of the dataset present significant challenges for the remote sensing object detection task.

Implementation details: We utilized the Faster R-CNN (Ren et al. 2015), like previous SSL works, and assembled it with our pre-trained remote sensing encoder. We employed the official implementation from Pytorch and adjusted several parameters to better accommodate the characteristics of object detection in remote sensing scenarios. We set anchor sizes to a range of 4 to 512 to capture more tiny objects like cars and ships, and set aspect ratios to a range of 0.2 to 5.0 to fit some slim targets. The RPN NMS threshold was set to 0.8, box score threshold was set to 0.01, and the detections for images were set to 200. All the images are resized to a size of 512 × 512 online, and the bounding boxes are also rescaled according to the images. We finetuned and evaluated our model with the AdamW optimizer for 100 epochs on the official splits. The batch size was set to 16, and the weight decay was set to 1e-3. The learning rate was set to 1e-4, and after warming up, it decreased according to the cosine scheduler. We set the drop path rate to 0.3 to enhance the model's generalization ability. For data augmentations, we used random flipping and color distortion.

Results and analyses: We finetuned our pre-trained RS foundation model on the DIOR dataset to evaluate the object detection ability of our method and report the mAP50 metric. As shown in Table 7, our method achieved 75.71% mAP50 on the DIOR dataset. Although this result surpasses the majority of existing remote sensing pre-training methods, there remains a performance gap compared to the current state-of-the-art. This discrepancy may be attributed to the global attention mechanism of the Vision Transformer, which, despite the introduction of spatial information via the ViT-Adapter, still lacks the inherent spatial priors present in architectures such as the Swin Transformer. Future work could further investigate and refine this aspect to enhance performance.

Table 7 Comparisons with other RS foundation models on the DIOR dataset.

| Method | Arch. | $mAP_{50}$ (%) |
|---|---|---|
| RVSA (2022) | ViTAE-B | 73.22 |
| Scale-MAE (2023) | ViT-L | 73.81 |
| SatMAE (2022) | ViT-L | 70.89 |
| TOV (2023) | ResNet-50 | 70.16 |
| CMID (2023) | Swin-B | 75.11 |
| GASSL (2021) | ResNet-50 | 67.40 |
| SSL4EO (2023) | ViT-S | 64.82 |
| CACo (2023) | ResNet-50 | 66.91 |
| SatLas (2023) | Swin-B | 74.10 |
| GFM (2023) | ViT-L | 72.84 |
| Skysense (2024) | ViT-L + Swin-H | 78.73 |
| **PerA** | ViT-G/16-1024 | 75.71 |

### *4.3. Comparative results*

#### *4.3.1. Comparison on uncurated data*

A series of comparative experiments were conducted on uncurated data. As depicted in Table 8, we pre-trained ViT-B with PerA and the DINO V2 method on the sampled pre-training dataset TOV-100k and RSRSD-100k. Unlike the curated TOV dataset, the images in the RSRSD dataset are not restricted to specific target locations, are closer to the characteristics of the data in practical interpreting tasks, and are easier to access. The PerA method extracts precise and concise semantics from sparse inputs, which enable the model to be pre-trained in arbitrarily cropped remote sensing images avoiding semantic inconsistency. We performed scene classification tasks on the AID dataset and report OA of test set to verify our conclusion. The experimental results indicate that, under identical experimental conditions, our method PerA obtained similar performance on both curated and uncurated datasets, whereas the baseline method DINO V2 exhibited significant performance degradation when applied to uncurated data. These experimental results demonstrate the superior robustness of our method and its enhanced capability in conducting pre-training with uncurated data. Furthermore, the experiments indirectly validate our method's effectiveness in reducing semantic inconsistency issues, as uncurated data presents a higher probability of semantic inconsistency occurrences due to scattered distribution. This remarkable adaptability to uncurated data can be attributed to the specially designed pretext task in our method, which enhances the semantic alignment of sample pairs for scattered geospatial objects while reducing the model's reliance on the spatial distribution of objects in the data.

Table 8 Comparison with baseline method on uncurated data.

| Method | Data | OA (%) |
|---|---|---|
| PerA | TOV-100k | 90.1 |
| | RSRSD-100k | 90.0 |
| DINO V2 | TOV-100k | 82.3 |
| | RSRSD-100k | 79.3 |

#### *4.3.2. Comparison with other SSL methods*

We evaluate semantic learning capability between different SSL methods in same setting. All experiments were performed using the ViT-B backbone with different approaches, including both CL and MIM. We pre-trained all these models on TOV-100k for 200 epochs and conducted downstream experiments on the AID dataset. As shown in Table 9, we report overall accuracy of AID dataset. Under the same experimental setting, our method achieves the best performance among a number of advanced SSL methods including CL and MIM. The results demonstrate the potential of our pre-training method for semantic understanding of remote sensing images and prove the effectiveness of combination CL and MIM methods.

Furthermore, we conduct comparisons of computational resource between different CL methods which have a similar structure. All the models are pre-trained with ViT-B on TOV-100k dataset. The images in dataset were cropped into 512 × 512, and batch size was set to 16. Resource usage information such as computational complexity and memory consumption is reported in Table 10. The result shows that, comparing with other CL methods, our method provides dramatic improvements in training speed, computational complexity, and memory consumption. Thanks to the reduction in redundant information from the sparse input, our method reduces image processing time to about half, and memory usage to approximately one-quarter, compared to the baseline approach DINO V2. The results indicate that our approach enables pretraining with lower resource consumption while still achieving strong performance.

Table 9. Comparison of the performance of different SSL methods under the same settings.

| Method | Type | OA (%) |
|---|---|---|
| DINO | CL | 80.1 |
| MAE | MIM | 84.9 |
| MoCo V3 | CL | 69.2 |
| DINOv2 | CL | 82.3 |
| PerA | CL | 90.1 |

Table 10. Comparison of the resource usage of different SSL methods under the same settings.

| Method | Throughput (image/s) | GFLOPs | Epoch time (min) | GPU memory (GB) |
|---|---|---|---|---|
| DINO | 33.90 | 1177.58 | 49.25 | 24.82 |
| MoCo V3 | 50.72 | 351.05 | 32.92 | 11.55 |
| DINO V2 | 69.28 | 1249.93 | 24.93 | 26.16 |
| PerA | 112.69 | 146.85 | 14.82 | 6.64 |

### *4.4. Ablation studies*

#### *4.4.1. Performance improvement*

Based on the DINO V2 framework, PerA implemented a series of performance optimizations and component improvements. These performance improvements are well demonstrated under combinations of the components. By introducing components progressively, we conducted a comparative analysis of our method's performance on the AID dataset. As shown in Table 11, we pre-trained models integrated from different components on TOV-100k for 200 epochs. Initially, we discarded IBOT patch-level alignment since the asymmetric design of our method learns between sample pairs with no shared regions, which would lead to IBOT failure. After applying spatial alignment, the model's performance was diminished due to spatial correspondence issues, as overly simplistic samples hinder its ability to learn high-level semantic features. However, when disjoint masks were introduced, the challenging pretext tasks helped restore the accuracy to normal levels — or even improve it further. Building on this, the pixel prediction significantly improved the semantic capabilities of the model, resulting in an overall accuracy of 90.1%. Additionally, we evaluate our approach without spatial alignment, which indicates that the improvements are the result of the combined effects of spatial alignment, sparse input, and pixel prediction.

Table 11. Ablation studies of the different improvements with ViT-B. SA: spatial alignment; DM: disjoint mask; PP: pixel prediction.

| Method | IBOT | SA | DM | PP | OA (%) |
|---|---|---|---|---|---|
| DINOv2 | √ | | | | 81.4 |
| PerA | | √ | | | 75.7 |
| | | √ | √ | | 84.9 |
| | | | √ | √ | 89.6 |
| | | √ | √ | √ | 90.1 |

***4.4.2. Ablation studies of different patch sizes***

According to the theory behind our method, using the smallest possible patch size can eliminate semantic inconsistency. However, the actual impact of patch size in experiments is more complex. Extremely small patch sizes significantly increase computational cost while also reducing the difficulty of the pre-text task, leading to performance degradation.

Several experiments were conducted with different patch sizes to evaluate the impact of patch size on model performance. We pre-trained models on TOV-100k for 200 epochs. As shown in Table 12, all these experiments were in the same settings, and we report the differences in results under different patch size configurations. We can observe that the model's performance improves when the patch size is reduced from 16 to 14, which is consistent with the general behavior of ViT. However, when the patch size was set to 8, potential agent task simplification leads to a noticeable decay in model performance.

Table 12. Performance on the AID dataset for different patch sizes.

| Method | Arch. | Patch size | OA (%) |
|---|---|---|---|
| PerA | ViT-B | 16 | 90.1 |
| | | 14 | 90.5 |
| | | 8 | 89.3 |

***4.4.3. Ablation studies on different mask ratios***

We conducted more detailed experiments to find the best ratios of disjoint masks, which are presented in Table 13. The disjoint masks divide each input remote sensing image into three parts: the student part ($s$), the learnable part ($l$), and the teacher part ($t$). The ratio of disjoint masks is a sensitive hyperparameter in our method. Through extensive experiments, we found that the pre-trained model performs optimally when the $t$ ratio is set to 50% or 60%. The pre-training performs best when the ratio between the $s$ part and the $l$ part is approximately 1:1. Due to the ground truth information interaction during the training, although the pixel information in the $l$ part is masked, it can still be implicitly captured through the reconstruction target.

Table 13. Ablation studies of different mask ratios.

| Method | $s$ ratio | $l$ ratio | $t$ ratio | OA (%) |
|---|---|---|---|---|
| PerA | 30% | 20% | 50% | 91.4 |
| | 20% | 30% | 50% | 91.9 |
| | 20% | 20% | 60% | 91.9 |
| | 30% | 10% | 60% | 90.7 |

#### *4.5. Visual analysis*

We conducted several visual analysis experiments to interpret the improvements introduced by our method and to demonstrate its advantages.

#### *4.5.1. Attention maps and reconstruction*

We performed several visualizations of the pre-trained model and identified that the model pre-trained by our method shares the specific properties of both CL and MIM.

Due to the unique architecture that combines the CL method with the MIM method, PerA possesses distinctive properties that are not found in other methods. As shown in Figure 4 (a), similar to the CL method DINO, the encoder pre-trained with PerA produces semantic-rich attention maps that clearly delineate the boundaries of objects in remote sensing images. In each image pair, the left image is the original image from the AID dataset, while the right image shows the attention map obtained from a pre-trained PerA encoder. The attention maps are generated from the multi-head attention module of ViT, where each attention head can capture different semantic features. Sometimes, the objects in these maps are so precise that even comparable to those produced by segmentation methods. Since the AID dataset was not involved during pre-training, these attention maps are obtained under a fully zero-shot situation, which demonstrates the strong semantic understanding ability of our method based on perfectly aligned sample pairs.

Moreover, due to the incorporation of the pixel reconstruction mechanism, the model pre-trained by PerA also exhibits powerful pixel reconstruction capability, which is usually associated with MIM methods. As illustrated in Figure 4 (b), the pre-trained encoder with its prediction head can accurately reconstruct heavily masked input remote sensing images. For each triplet of images, the left image is the masked input, the center image is the reconstructed output, and the right image is the original remote sensing image. These pixel reconstruction predictions are also conducted on the AID dataset, which was not involved during pre-training. The fact that the pre-trained model can almost perfectly reconstruct images with up to 70% masking, without any prior knowledge of the dataset, further demonstrates its deep understanding of both image structure and semantics. This shows that the model can understand sparse input information and transfer it into essential, semantic-rich features, while maintaining strong robustness to noise.

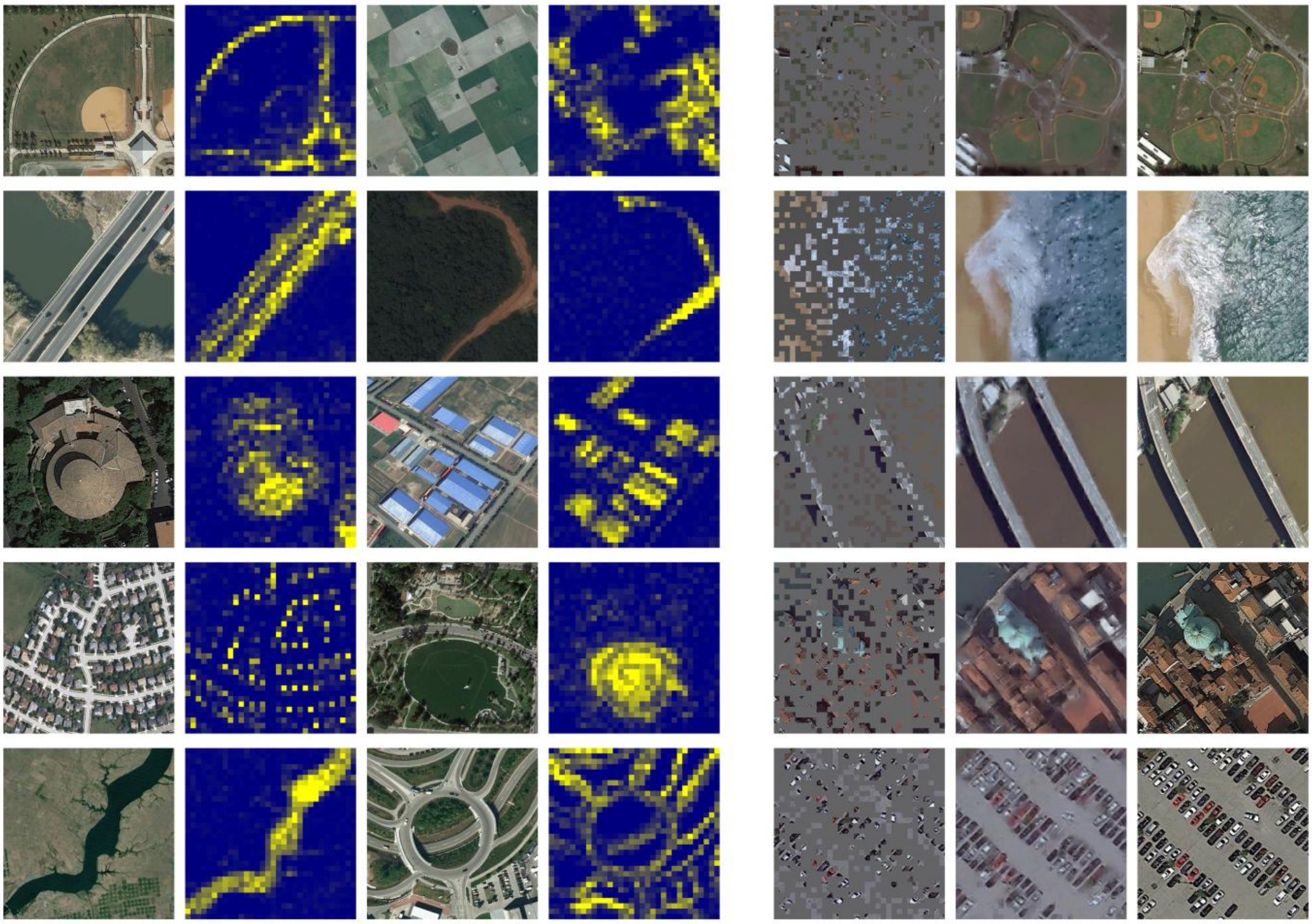

(a) (b)

Figure 4. Visualization of self-attention maps and reconstruction results on the AID dataset (16 × 16 patch size). Note that all samples were unseen during pre-training. (a) Self-attention visualization: original image (left) and attention map (right); (b) Reconstruction visualization: masked image (left), reconstructed output (middle), and ground-truth (right).

#### *4.5.2. t-SNE visualization*

We conducted t-SNE visualization of the feature distributions produced by the baseline method DINO V2 and the improved method PerA on the AID dataset, clearly demonstrating the improvement in semantic representation achieved by our approach. As presented in Figure 5, all classified features are downscaled and visualized as scatter plots in feature space. In the plots, each color represents a different category. It can be observed that our method achieves significant improvement over the baseline in teams of intra-class compactness, with a noticeable reduction in misclassified feature points within each class. This intuitively reflects the improved classification capability of the proposed method on downstream tasks.

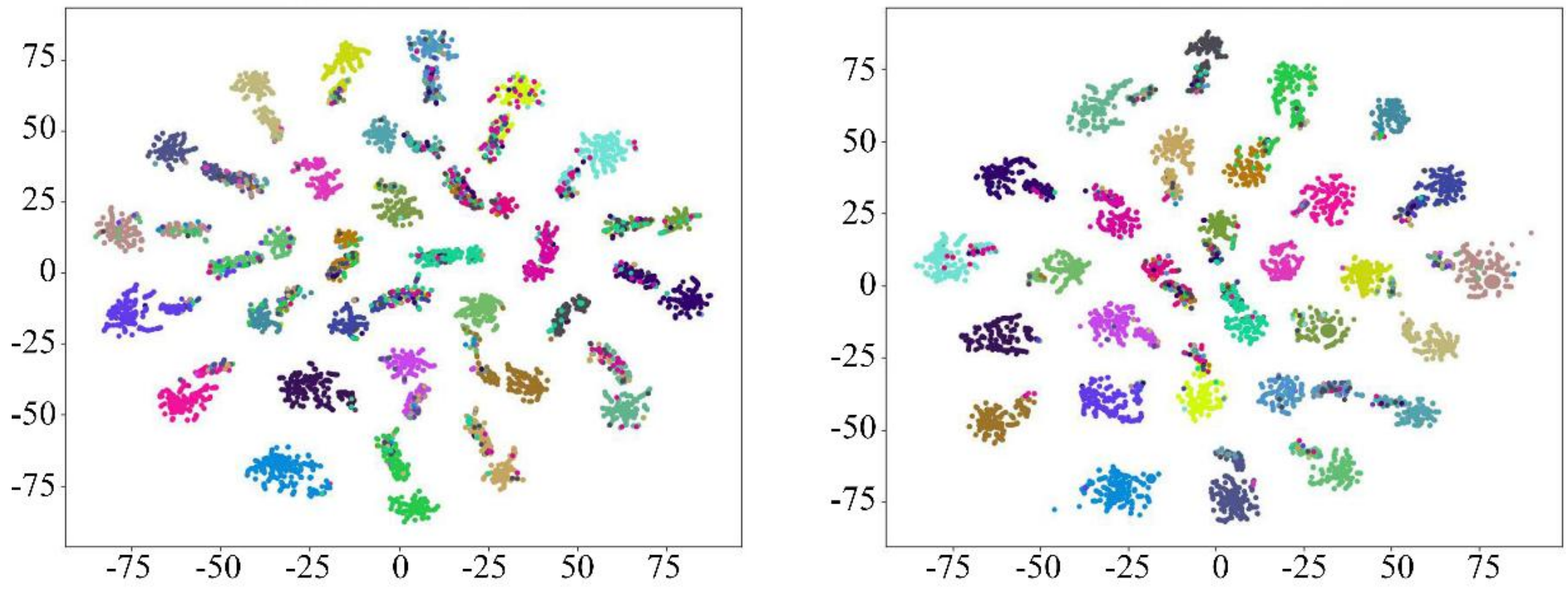

Figure 5. Comparison of t-SNE maps of (a) DINOv2 and (b) PerA on the AID dataset.

#### *4.5.3. Feature histogram*

To further analyze what is the difference in representations between our improved method and baseline, we conducted histogram visualizations for pre-trained encoder feature outputs. As depicted in Figure 6 (a), We compared the distributions of output feature differences between the student and teacher encoders for our improved method and the baseline DINO V2. The histograms reflect a noticeable increase in the frequency of lower difference values and a corresponding decrease in the frequency of higher values. These statistical results indicate that the improved method leads to significantly reduced feature differences between the teacher and student networks and a better high-level semantic understanding.

Additionally, as shown in Figure 6 (b), we compared output feature distributions of the pre-trained model between DINO V2 and PerA. In the histograms of feature distribution, our method exhibited a more pronounced peak and a significant reduction in the discrete values. The contraction of the feature range, accompanied by improved downstream task performance, suggests that the model has learned more generalizable and high-level features from sparse inputs, effectively discarding redundant information.

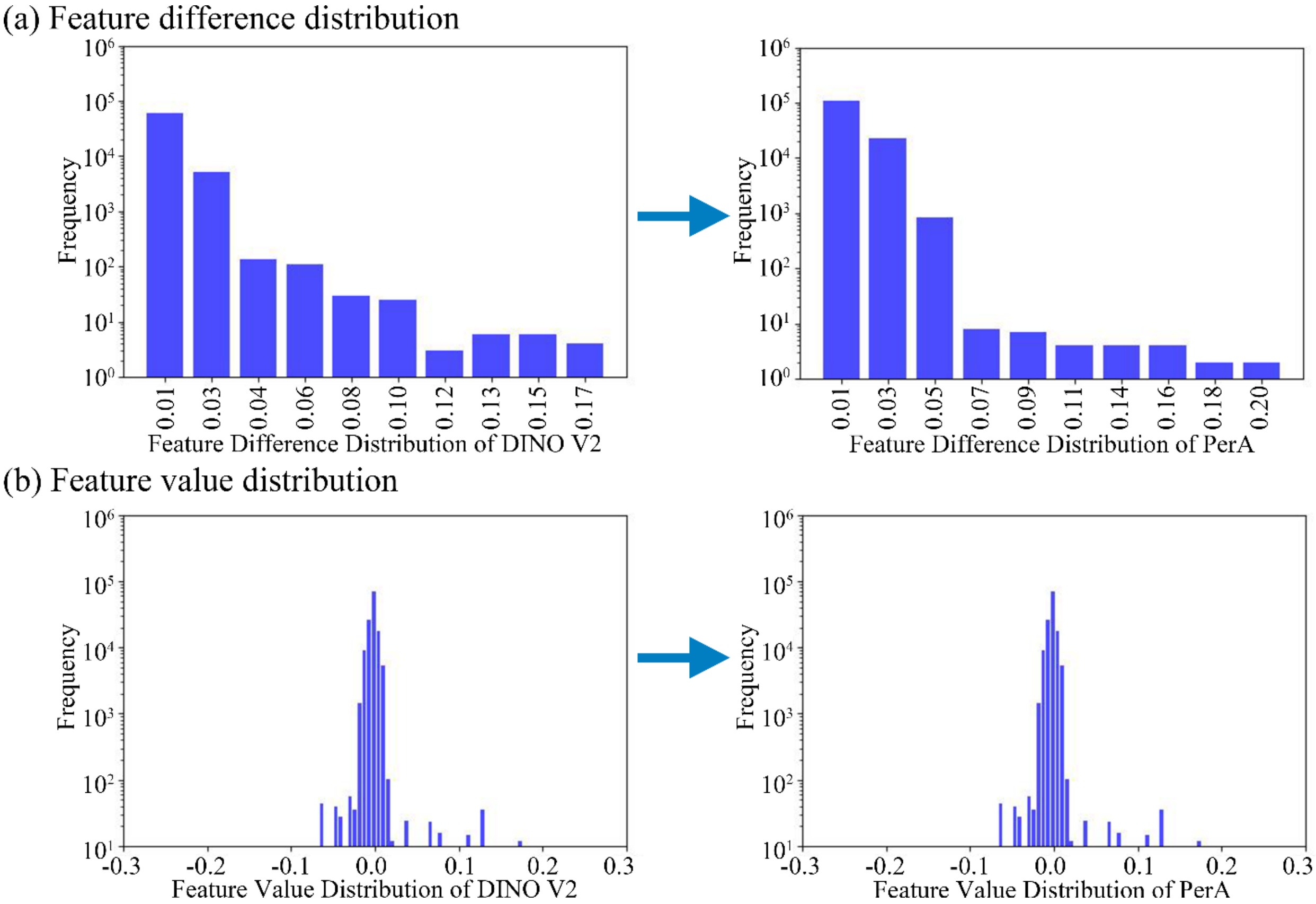

Figure 6. Analysis of feature distribution between DINO V2 and PerA: (a) Comparison of output feature difference distributions between student and teacher; (b) Comparison of the output feature value distribution.

## 5. Conclusions

In this paper, we propose a contrastive learning pre-training method named PerA, which is designed to adapt remote sensing images. PerA is intended to generate sample pairs with as consistent semantics as possible while maximizing the utilization of the samples. Compared with other remote sensing unlabeled pre-training models, our method obtained remarkable performance on several downstream tasks with a limited model scale. We also present an unlabeled remote sensing pre-training dataset called RSRSD-5m, which contains about 5 million remote sensing images. To the best of our knowledge, this is one of the largest publicly available unlabeled remote sensing datasets. In the future, we aim to reduce the cost of using models through methods such as knowledge distillation, enabling their better application in practical remote sensing monitoring tasks.

## Disclosure statement

No potential conflict of interest was reported by the author(s).

## Funding

This work is supported by the National Key R&D Program of China [Grant number 2023YFB3907600], the Chinese Academy of Surveying and Mapping Fundamental Research Funds [Grant number AR2420], and the Zhejiang Province "Vanguard" and "Geese Leading" Research and Development Plan [Grant number 2025C01073].

**Data availability statement**

The source code, weights, and datasets are available at https://github.com/SathShen/PerA.

**Notes on contributors**

***Hengtong Shen*** received the M.E. degree in Photogrammetry and Remote Sensing from the Chinese Academy of Surveying and Mapping in 2025. He is currently pursuing a Ph.D. degree at Wuhan University. His research interests include intelligent remote sensing interpretation, self-supervised learning for remote sensing, and deep learning.

***Haiyan Gu*** received the Ph.D. degree in Photogrammetry and Remote Sensing from Wuhan University in 2015. She is currently a researcher in Chinese Academy of Surveying and Mapping. Her research direction is intelligent interpretation and high-performance processing of high-resolution remote sensing images, and she is committed to solving technical problems in practical business.

***Haitao Li*** received the Ph.D. degree in pattern recognition and intelligent systems from Huazhong University of Science and Technology. He is currently a Researcher with the Chinese Academy of Surveying and Mapping. His research interests lie at the intersection of remote sensing image processing, computer vision, and artificial intelligence, with a specific focus on intelligent interpretation and high-performance computing.

***Yi Yang*** received the B.E. and M.E. degree in Huazhong University of Science and Technology in 2009 and 2012, respectively. He is currently an associate researcher with the Natural Resources Investigation and Monitoring Research Center, Chinese Academy of Surveying and Mapping, Beijing. His research interests include remote sensing image processing and deep learning.

***Agen Qiu*** received the Ph.D. degree in cartography and geographical information engineering from Wuhan University in 2017. He is currently a researcher at the Chinese Academy of Surveying and Mapping. His research interests include spatio-temporal big data analysis and mining, and intelligent modeling of geographic information.

## ORCID

Hengtong Shen https://orcid.org/0009-0002-8190-4927

Haiyan Gu https://orcid.org/0000-0002-5934-9579

Haitao Li https://orcid.org/0009-0007-7595-8001

Yi Yang https://orcid.org/0009-0006-2715-1177

Agen Qiu https://orcid.org/0009-0007-2195-9252